\documentclass[10pt]{article}
\usepackage{url}
\usepackage{soul}

\usepackage{graphicx}
\usepackage{epstopdf}
\usepackage{amssymb}
\usepackage{amsmath}
\DeclareMathOperator{\sech}{sech}
\usepackage{float} % For placing figures
\usepackage{wrapfig}
\usepackage{natbib}
\usepackage[ruled,vlined]{algorithm2e}
\SetKwProg{Timestep}{Time Step}{}{end Time Step}
\SetKwProg{Momentfitting}{Moment Fitting}{}{end Moment Fitting}
\SetKwProg{Boundarymomentfitting}{Boundary Moment Fitting}{}{end Boundary Moment Fitting}

\usepackage{geometry}
\usepackage[normalem]{ulem}

\renewcommand{\vector}[1]{\mathbf{#1}}    % vector notation
\newcommand\extrafootertext[1]{%
    \bgroup
    \renewcommand\thefootnote{\fnsymbol{footnote}}%
    \renewcommand\thempfootnote{\fnsymbol{mpfootnote}}%
    \footnotetext[0]{#1}%
    \egroup
}

\usepackage[hidelinks]{hyperref}

\begin{document}
\sloppy
%\justifying

\title{\Large\bf An Energy Conserving Vlasov Solver That Tolerates \\Coarse Velocity Space Resolutions:\\
Simulation of MMS Reconnection Events}

\author{F. Allmann-Rahn$^1$, S. Lautenbach$^2$, R. Grauer$^1$ \\
        \small $^1$ Institute for Theoretical Physics I, Ruhr-University Bochum, Germany \\
        \small $^2$ Department of Physics, University of Alberta, Edmonton, Alberta T6G 2E1, Canada
        }

% \correspondingauthor{Florian Allmann-Rahn}{far@tp1.rub.de}

% \begin{itemize}
% \item A moment fitting continuum Vlasov-Maxwell solver is presented that preserves positivity of the distribution function
% and conserves total energy.
% \item The method behaves well at low velocity space resolutions, making it competitive with particle-in-cell (PIC) methods
% concerning computational cost.
% \item There is good agreement of the simulations with measurements of magnetic reconnection by the MMS spacecraft.
% \end{itemize}

\date{}
\maketitle

\begin{abstract}

Vlasov solvers that operate on a phase-space grid are highly accurate
but also numerically demanding. Coarse velocity space resolutions, which
are unproblematic in particle-in-cell (PIC) simulations, can lead to
numerical heating or oscillations in standard continuum Vlasov methods.
We present a new dual Vlasov solver which is based on an established positivity preserving
advection scheme for the update of the distribution function and an energy
conserving partial differential equation solver for the kinetic update of mean
velocity and temperature. The solvers work together via moment fitting
during which the maximum entropy part of the distribution function is
replaced by the solution from the partial differential equation solver. This numerical scheme
makes continuum Vlasov methods competitive with PIC methods concerning
computational cost and enables us to model large scale
reconnection in Earth's magnetosphere with a fully kinetic continuum
method. The simulation results agree well with measurements by the MMS spacecraft.

\end{abstract}

\extrafootertext{Corresponding author: Florian Allmann-Rahn, \texttt{far@tp1.rub.de}}

\section{Introduction}\label{sec:introduction}

In many plasmas, for example the space plasma around the Earth and the sun and plasmas
in fusion devices, collisions are rare. Therefore kinetic methods are necessary
to accurately model these plasmas. Kinetic continuum Vlasov simulations provide
an accurate and noise-free representation of velocity space,
but solve the Vlasov equation on a phase space grid which is numerically challenging.
In order to avoid unphysical negative values for the particle distribution function,
positivity preserving limiters can be introduced. These, however,
lead to numerical heating of the plasma so that conservation of total energy is
violated. In Vlasov solvers that conserve energy, on the other hand,
special care must be taken to keep numerical oscillations (caused by negative
values of the distribution function) under control.
Otherwise, their usability over longer time-spans is limited
in simulations with prominent non-linear effects. Both numerical heating
and non-positivity become more problematic at low resolutions/large
cell sizes in velocity space. Published full Vlasov
simulations of magnetic reconnection 
(e.g. \citet{schmitz-grauer:2006, pezzi-cozzani-califano-etal:2019, pezzi-liang-juno-etal:2021,
liu-cai-lapenta-etal:2021}) did not go beyond the GEM reconnection setup \citep{birn-drake-shay-etal:2001}
which is computationally manageable due to the small system size.

To give an overview, we want to mention just a few of the many available schemes for
solving the Vlasov equation on a phase-space grid, together with related implementations in high-performance
computing codes. The pioneer work of \citet{cheng-knorr:1976} used a semi-Lagrangian
approach with spline or Fourier interpolation. A high-performance implementation that
uses Lagrange interpolation can be found in \citet{kormann-reuter-rampp:2019}.
Such semi-Lagrangian solvers are still most popular due to their computational
efficiency. Here, we use the third-order semi-Lagrangian positive and flux-conservative
(PFC) scheme in \citet{filbet-sonnendruecker-etal:2001}
which ensures positivity of the distribution function and conserves mass and fluxes. A similar
approach with higher-order accuracy can be found in \citet{tanaka-yoshikawa-minoshima-etal:2017}.
The PFC scheme was first applied to magnetic reconnection in \citet{schmitz-grauer:2006}.
More recently, discontinuous Galerkin methods were adopted for Vlasov simulations e.g.~a
semi-Lagrangian approach in \citet{rossmanith-seal:2011}. A discontinuous Galerkin scheme
with Runge-Kutta time integration, implemented for the full Vlasov-Maxwell system, is
given in \citet{juno-hakim-etal:2018} and \citet{hakim-juno:2020}. This scheme
conserves total energy, and although it does not preserve positivity of the distribution function,
the plasma turbulence simulation in \citet{juno-hakim-etal:2018} was not
impaired by numerical oscillations.

Magnetic reconnection is a fundamental energy conversion process in plasmas
throughout the universe and can excite instabilities in fusion devices.
In 2015 the Magnetospheric Multiscale Mission (MMS) spacecraft was launched
to directly measure reconnection in the Earth's magnetosphere
\citep{burch-moore-torbert-etal:2016}.
There have been successful comparisons between simulations of reconnection
using fully kinetic particle-in-cell (PIC) models and MMS measurements
\citep{nakamura-genestreti-etal:2018,liu-lu-turner-etal:2020,Lu-Wang-Lu-etal:2020}.
In the particle-in-cell (PIC) method
the velocity space is represented by super-particles, which represent a large amount of
actual particles, using a Monte-Carlo approach.
This is very efficient because the method is stable and produces
reasonable results even at low velocity space resolution (i.e.~low numbers of
particles) and in many cases the number of particles is higher in regions
of interest so that the simulation accuracy adapts nicely to the physical configuration.
However, when too few particles are used, discrete particle noise can limit
simulation accuracy \citep{nevins-hammett-dimits-etal:2005,juno-swisdak-tenbarge-etal:2020}.
Both the continuum Vlasov method and the PIC method have their respective strengths,
but the primary reason that continuum solvers could so far not compete
with PIC is the numerical difficulty of treating low velocity space resolutions.

In this paper we present a method to make a Vlasov solver both positivity preserving
and energy conserving by means of moment fitting. This relaxes the numerical necessity of
high velocity space resolutions and thus enables us to address large-scale problems with
a fully kinetic continuum solver. We show in
comparisons between reconnection simulations and MMS measurements that
the Vlasov method can provide an accurate representation of the electric
field at moderate computational cost.

\section{Physical and Numerical Models}

\subsection{The Vlasov-Maxwell System and its Numerical Representation}\label{sec:vlasov_maxwell}

A collisionless plasma evolves according to the Vlasov equation
\begin{equation}
\frac{\partial f_{s}}{\partial t} +\mathbf{v} \cdot \nabla f_{s} + \frac{q_{s}}{m_{s}} (\mathbf{E + v \times B}) \cdot \nabla_{v} f_{s} = 0,
\label{eq:vlasov} \end{equation}
where ${f_s(\mathbf{x},\mathbf{v},t)}$ is the particle distribution function for each species $s$.
A collision operator can be added to the right hand side of the equation in the case of collisional plasmas.
From the distribution function physical quantities can be obtained by taking moments. The particle density is given by
${n_s(\mathbf{x}, t) = \int f_s(\mathbf{x}, \mathbf{v}, t) \text{d}\mathbf{v}}$ and the mean velocity is
${\mathbf{u}_s(\mathbf{x},t) = \frac{1}{n_s(\mathbf{x}, t)} \int \mathbf{v} f_s(\mathbf{x}, \mathbf{v}, t) \text{d}\mathbf{v}}$.
The second and third moment (multiplied by mass) are
momentum flux density ${\mathcal{P}_{s} =  m_{s} \int \mathbf{v} \otimes \mathbf{v} f_{s} \text{d}\mathbf{v}}$ and
energy flux density ${\mathcal{Q}_{s} =  m_{s} \int \mathbf{v} \otimes \mathbf{v} \otimes \mathbf{v} f_{s} \text{d}\mathbf{v}}$, respectively,
where $\otimes$ denotes the tensor (outer) product.
Heat flux ${\mathrm{Q}}$ is related to the third moment ${\mathcal{Q}}$ like
${\mathrm{Q}_{ijk} = \mathcal{Q}_{ijk} - \text{sym}(\vector u \otimes \mathcal{P})_{ijk} + 2 m n u_i u_j u_k}$
and temperature ${\mathrm{T}}$ to the second moment ${\mathcal{P}}$ like
${\mathrm{T}_{ij} = (\mathcal{P}_{ij} - m n u_i u_j)/(n k_B)}$. From the heat flux tensor the heat flux vector
can be obtained as ${\mathbf{q}_i = \sum_j \mathrm{Q}_{ijj}}$.

Taking moments of the complete Vlasov equation, a set of fluid equations follows:
\begin{equation}
\frac{\partial n_{s}}{\partial t} + \nabla \cdot (n_{s} \mathbf{u}_{s}) = 0 \, ,
\label{eq:tenmoment_continuity} \end{equation}
\begin{equation}
m_{s} \frac{\partial (n_{s} \mathbf{u}_{s}) }{\partial t} =
n_{s} q_{s} (\mathbf{E} + \mathbf{u}_{s} \times \mathbf{B}) - \nabla \cdot \mathcal{P}_{s} \, ,
\label{eq:tenmoment_movement} \end{equation}
\begin{equation}
    \frac{\partial \mathcal{P}_{s}}{\partial t} - q_{s} (n_{s} \text{sym}(\mathbf{u}_s \otimes \mathbf{E})
    + \frac{1}{m_{s}} \text{sym}(\mathcal{P}_{s} \times \mathbf{B}))
    = - \nabla \cdot \mathcal{Q}_{s} \, .
\label{eq:tenmoment_energy} \end{equation}
Here, ${\text{sym}}$ is the symmetrization, i.e.~the sum over permutations of indices
to make the tensors symmetric, for example
${\text{sym}(\mathbf{u} \otimes \mathbf{E})_{ij} = u_i E_j + u_j E_i}$,
and the vector product $\times$ is generalized to tensors.

The fluid equations are exact but contain more unknowns than equations, in particular there is
no equation for ${\mathcal{Q}_{s}}$. In the moment fitting Vlasov method we obtain ${\mathcal{Q}_{s}}$
directly from the distribution function so that the fluid equations are equivalent with the
Vlasov equation. Multi-fluid methods instead make an approximation with a physically motivated
closure expression as described in Sec.~\ref{sec:fluid_solver}.

The evolution of electric and magnetic fields is determined by Maxwell's equations
  \begin{displaymath}
  \nabla \cdot \mathbf{E} = \frac{\rho}{\epsilon_{0}}\,,\quad
  \nabla \cdot \mathbf{B} = 0\,,\quad
  \nabla \times \mathbf{E} = - \frac{\partial \mathbf{B}}{\partial t}\quad\text{and}\quad
  \nabla \times \mathbf{B} = \mu_{0} \mathbf{j} + \mu_{0} \epsilon_{0} \frac{\partial \mathbf{E}}{\partial t}\,.
  \end{displaymath}
Together with the Vlasov equation they form the Vlasov-Maxwell system of equations which fully describes
the plasma dynamics.

In the electromagnetic simulations we normalize length over ion inertial length ${d_{i,0}}$ based on
density ${n_0}$, velocity over ion Alfv\'{e}n velocity ${v_{A,0}}$ based
on the magnetic field ${B_0}$, time over the inverse of the ion
cyclotron frequency ${\Omega_{i,0}^{-1}}$, mass over ion mass ${m_i}$,
electric charge in ion charge ${q_i}$,
and vacuum permeability ${\mu_0 = 1}$ as well as Boltzmann constant ${k_B = 1}$.

The Vlasov equation is solved by means of the
PFC method in \citet{filbet-sonnendruecker-etal:2001}
and the velocity splitting is realized via the backsubstitution method
\citep{schmitz-grauer:2006b}. We use zero-flux boundary conditions in
velocity space to ensure conservation of particle density.
The fluid solver utilizes a centrally weighted essentially
non-oscillating (CWENO) method \citep{kurganov-levy:2000} and the
third-order Runge-Kutta scheme in \citet{shu-osher:1988}.
The finite-difference time-domain (FDTD) method is employed for
the Maxwell equations.

The schemes were implemented in the \textit{muphy2} multiphysics plasma simulation code
developed at the Institute for Theoretical Physics I, Ruhr University Bochum.
The framework part of the code is written in C++ whereas the pure computational
parts are written in Fortran to benefit from its excellent performance when dealing with
multi-dimensional arrays. Parallelization is done via domain decomposition and MPI.
All solvers are fully ported to GPUs with highly optimized OpenACC. That
way the same code base can be used for both GPU and CPU computations.

\subsection{Moment Fitting for a Positive and Energy Conserving Vlasov Solver}\label{sec:moment_fitting}

As discussed, numerical solutions to the Vlasov equation \eqref{eq:vlasov} are not
positivity preserving and energy conserving at the same time, whereas the fluid equations
\eqref{eq:tenmoment_continuity}--\eqref{eq:tenmoment_energy} are unproblematic
in this regard. The idea for getting a both positive and energy conservative
Vlasov solver is therefore to obtain the heat flux moment from the distribution
function and use it in the fluid equations to get an exact kinetic solution
to the Vlasov equation for the momentum and energy moments from the fluid solver.
These moments are then used to update the maximum entropy part
(in the Boltzmann sense) of the distribution function so that
energy is conserved. We call this method \textit{moment fitting}. The update
of the distribution function is realized by calculating the ten-moment Maxwellian
part of the distribution function and replacing it with the ten-moment Maxwellian
calculated from the exact fluid solver moments. As density is conserved
in both methods, this operation conserves the distribution function. Furthermore, it
affects not more than the originally not conserved momentum and energy
for which we now have numerically more accurate as well as kinetically correct solutions.

The original idea of moment fitting for improving Vlasov solvers is from
\citet{trost-lautenbach-grauer:2017} and a simple form had been used before
in \citet{rieke-trost-grauer:2015} for spatial coupling of Vlasov and
five-moment fluid models. \citet{trost-lautenbach-grauer:2017} adapted
moments by shifting, stretching and rotating the distribution function.
While that is certainly a good and successful approach, we want to address
two subtleties that are involved. First, in that method the
equation for determining the adaption of the distribution function
is underdetermined and needs to be solved with a optimization algorithm
under the additional constraint that the adaption is as close to unity
as possible. Second, the whole distribution function is adapted and not
only the part directly related to the maximum entropy ten-moment solution.
That means also higher moments, which the fluid solver gives no kinetic
solution for, are changed on basis of the new momentum and temperature.
While this may often be desired, it is not clear that the
result is in any case still a valid solution to the Vlasov equation.
With the new method of exchanging only what the fluid solver
gives definite answers to, we can be sure to not change the kinetic physics.

\LinesNumbered
\begin{algorithm}\label{alg:timestep}
\SetAlgoLined
 \Timestep{}{
  Half step of Maxwell solver\\
  Calculate third moment $\mathcal{Q}^{t}$ from $f^t$\\
  Full Vlasov Leapfrog Step to advance to $f^{t+1}$\\
  Calculate $\mathcal{Q}^{t+1}$ from $f^{t+1}$\\
  Interpolate to get $\mathcal{Q}^{t+1/2}$\\
  Full Runge-Kutta Fluid Step (input $\mathcal{Q}$ at appropriate times)\\
  \Momentfitting{}{
    Calculate moments $n_{\text{V}}$, $\mathbf{u}_{\text{V}}$, $\mathcal{P}_{\text{V}}$ from $f$\\
    Calculate ten-moment Maxwellian $f_{\text{M,V}}$ from $n_{\text{V}}$, $\mathbf{u}_{\text{V}}$, $\mathcal{P}_{\text{V}}$\\
    Rescale the fluid solver's moments $n_{\text{F}}$, $\mathbf{u}_{\text{F}}$, $\mathcal{P}_{\text{F}}$ by
        $n_{\text{V}}/n_{\text{F}}$ to ensure conservation of $f$\label{alg:timestep:rescale}\\
    Calculate ten-moment Maxwellian $f_{\text{M,F}}$ from the fluid solver's moments\\
    Exchange ten-moment Maxwellians: $f = f - f_{\text{M,V}} + f_{\text{M,F}}$\\
    Limit $f$ \label{alg:timestep:limit}
  }
  Second half step of Maxwell solver\\
 }
 \caption{Time stepping of the moment fitting Vlasov-Maxwell solver as it is implemented}
\end{algorithm}

A time step of the moment fitting Vlasov solver as we have implemented
it is shown in Algorithm~\ref{alg:timestep}.
Since we use a semi-Lagrangian solver
for the Vlasov equation and a Runge-Kutta solver for the Fluid equations,
the Vlasov step is done first so that the third moment that the fluid
solver needs is available at all Runge-Kutta times. Since the divergence
of the third moment is needed in Eq.~\eqref{eq:tenmoment_energy} it is
incorporated into the CWENO reconstruction process. Concerning the
moment fitting itself, we want to elaborate on some details.
The maximum entropy distribution function based on density $n$, mean velocity $\mathbf{u}$
and temperature tensor
${\mathrm{T} = \frac{m}{k_B n} \int (\mathbf{v-u}) \otimes (\mathbf{v-u}) f \text{d}\mathbf{v}}$,
which we here call the ten-moment Maxwellian, is given by
\begin{displaymath}
  f_M(\mathbf{x},\mathbf{v}) = n(\mathbf{x}) \left(\frac{m}{2 \pi k_B}\right)^{N/2} 
  \exp\left( -\frac{m}{2 k_B} (\mathbf{v}-\mathbf{u}(\mathbf{x}))^{\mathrm{t}}\ \mathrm{T}^{-1}\ 
  (\mathbf{v}-\mathbf{u}(\mathbf{x})) \right) / \sqrt{\det \mathrm{T}(\mathbf{x})},
\end{displaymath}
where $(\mathbf{v-u})^\mathrm{t}$ is the transpose of $\mathbf{v-u}$ and $N$ is the dimensionality
of velocity space. To have the distribution function conserved
when exchanging the ten-moment Maxwellians, the respective densities must be
identical. Both solvers conserve mass and solve the same equation
so in principal they should also yield the same density. There are,
however, small numerical errors which we compensate by rescaling the fluid
moments by the density obtained from the distribution function
(Alg.~\ref{alg:timestep}, line \ref{alg:timestep:rescale}).
That way we make sure that only unconserved quantities are changed during this
step. Especially where the distribution function is close to zero, the exchange of the ten-moment
Maxwellians can sometimes turn the value of the distribution function negative.
In the simulations in this paper, we limit $f$ by setting the respective values to zero in
these cases (line \ref{alg:timestep:limit}) followed by a rescaling of the distribution function to the
density before the exchange of the ten-moment Maxwellians.
This may not be very elegant, but the good energy conservation
properties discussed in the benchmark example sections justify the approach.
Overall, the presented moment fitting method makes as few changes as possible
to the distribution function when the full maximum entropy information
on momentum and energy from the fluid solver is kept.

If the fluid scheme did additionally conserve total momentum, as shown
e.g.~in \citet{balsara-amano-garain-etal:2016}, the moment fitting
would make the Vlasov solver conserve total momentum as well. So far
it does at least improve the momentum conservation. The published
momentum-conservative fluid solvers use five-moment models which
make the physical simplification that ${\mathrm{P}-\frac13\text{tr}(\mathrm{P})\,\text{id} = 0}$.
Thus, they are not exact and cannot be used in the moment
fitting method. The next step will be to transfer the ideas
of available momentum-conservative fluid solvers to the ten-moment model
which can then be utilized for a positivity preserving
and mass, momentum and energy conserving Vlasov solver.

\subsection{Fluid Solver with Gradient Heat Flux Closure}\label{sec:fluid_solver}
Ten-moment multi-fluid simulations close the hierarchy of equations with
a heat flux approximation. Here, we use a modified version of the temperature
gradient closure in \citet{allmann-rahn-trost-grauer:2018} and \citet{allmann-rahn-lautenbach-grauer-etal:2021}.
The closure is based on the one-dimensional Landau fluid closure
in \citet{hammett-perkins:1990} (also \citet{hammett-dorland-perkins:1992}) and takes inspiration
from \citet{wang-hakim-etal:2015}. Originally the gradient closure used
the gradient of the pressure's deviation from isotropy. Instead, we now simply
take the gradient of the temperature so that the heat flux approximation is
given by

\begin{equation}
\nabla \cdot \mathrm{Q}_{s} = -\frac{\chi}{k_{s,0}}\,n_s\,v_{t,s}\,\nabla^{2}\,\mathrm{T}_{s}\,.
\label{eq:gradient_closure}\end{equation}

We choose ${k_{s,0} = 1/d_s}$ as the typical spatial frequency.
The dimensionless parameter is ${\chi = 2 \sqrt{2/\pi}}$ as in \citet{hammett-perkins:1990}
and ${v_{t,s} = \sqrt{\frac{k_B\ \text{tr}(\mathrm{T}_s)/3}{m_s}}}$
is the thermal velocity.

A closure that is similar to the one we use here can be found in
\citet{ng-hakim-etal:2020}. The only difference is that they employed
a symmetrization in the process of generalizing the Landau fluid closure to
three dimensions. The closure given by Eq.~\eqref{eq:gradient_closure} yields in many cases
more accurate results than the original formulation in \citet{allmann-rahn-trost-grauer:2018}.

\section{Benchmark Problems}

\subsection{Landau Damping and Two-Stream Instability}\label{sec:landau_twostream}

From a theoretical point of view, it is clear that the combination of a
Vlasov and a fluid solver in the presented way yields correct solutions
to the Vlasov equation (within the accuracy of the solvers).
In this section we want to confirm at the example of
standard tests for kinetic plasma solvers that this is indeed the case also
numerically. In these test problems physical effects occur over short time spans and the resolution
is high so that the standard PFC scheme does conserve energy and momentum
precisely. That makes the setups ideal for verifying the moment fitting scheme
which should yield identical results. No correction of the moments is
to be expected but the Maxwellian part of the distribution function will still
be fully handled by the fluid solver with kinetic heat flux input.

\begin{figure}[h!]
\centering
\includegraphics[width=0.9\textwidth]{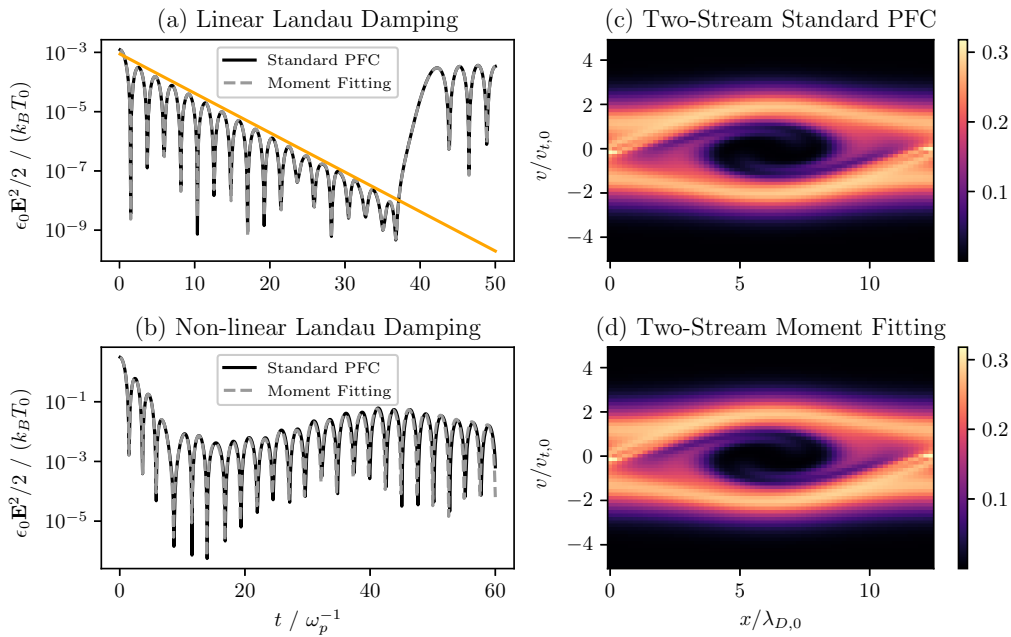}\\
\vspace{-0.5cm}
\caption{Comparison of the standard PFC and the moment fitting Vlasov solvers.
(a) Electric energy next to the analytical damping rate (orange line) in linear
Landau damping, (b) electric energy in non-linear Landau damping,
(c) and (d) distribution function $f_e$ obtained from the two methods in a two-stream instability setup
at $t = 30\,\omega_{p,0}^{-1}$.
}
\label{fig:landau_damping}\end{figure}

The discussed setups focus on electron effects and are purely electrostatic.
Therefore, we use a Poisson solver to calculate the electric field and choose
the following normalization: Time in inverse electron plasma frequency $\omega_{p,0}^{-1}$,
length in electron Debye length $\lambda_{D,0}$, velocity in electron thermal
velocity $v_{t,0}$, mass in electron mass $m_e$, temperature in initial electron temperature $T_0$
and finally ${\epsilon_0 = 1}$.
The initial condition for the Landau damping setup is
${f_{e,0}(x,v) = \frac{1}{\sqrt{2 \pi}}\,\exp(-v^2/2)\,(1 + \alpha \cos(k x))}$
with ${k = 0.5\,\lambda_{D,0}^{-1}}$ and a static and spatially uniform neutralizing ion background.
The one-dimensional domain has an extent of ${-2 \pi\,\lambda_{D,0}}$ to ${2 \pi\,\lambda_{D,0}}$ and
the velocity space of $-v_{\text{max}}$ to $v_{\text{max}}$. In the
linear Landau damping case it is ${v_{\text{max}} = 4.5\,v_{t,0}}$ and ${\alpha = 0.01}$
with a resolution of ${32 \times 32}$ cells,
whereas in the non-linear case it is ${v_{\text{max}} = 6\,v_{t,0}}$ and ${\alpha = 0.5}$
with a resolution of ${32 \times 64}$ cells.
The two-stream instability initial condition is 
${f_{e,0}(x,v) = \frac{1}{\sqrt{2 \pi}}\,v^2\,\exp(-v^2/2)\,(1 + \alpha \cos(k x))}$
with ${v_{\text{max}} = 5\,v_{t,0}}$, ${\alpha = 0.01}$ and a domain from
$0$ to ${4 \pi\,\lambda_{D,0}}$. The resolution is ${64 \times 64}$ cells.

Fig.~\ref{fig:landau_damping}(a) shows the damping of electric wave energy obtained
from standard PFC (black) and moment fitting (gray dotted) simulations of linear
Landau damping together with the analytic damping rate (orange).
The results are mostly identical and in both cases the damping
rate is ${\gamma = 0.153\,\omega_{p,0}^{-1}}$ (analytic: ${\gamma = 0.1533\,\omega_{p,0}^{-1}}$)
and the oscillation frequency is ${\omega = 1.41\,\omega_{p,0}}$
(analytic: ${\omega = 1.416\,\omega_{p,0}}$).
Non-linear Landau damping is also equally well represented by both methods as
shown in Fig.~\ref{fig:landau_damping}(b). The same results can be
found in \citet{filbet-sonnendruecker-etal:2001}. In
Fig.~\ref{fig:landau_damping}(c) and \ref{fig:landau_damping}(d) it is shown that the distribution
functions obtained from the standard PFC scheme and the moment fitting scheme
also match in the highly non-Maxwellian case of a two-stream instability.

\subsection{Orszag-Tang Turbulence}\label{sec:orszag_tang}
A plasma turbulence setup is well-suited for testing the moment fitting Vlasov
solver (and also the gradient fluid solver) because of the broad range of
plasma phenomena that are relevant for the dissipation of energy like multiple
types of waves, magnetic reconnection and Landau damping.
The Orszag-Tang turbulence setup we use has periodic
boundary conditions and thus is also a good setup to check the solvers' conservation
properties.
As we will show, the moment fitting Vlasov solver yields results that agree
with those from a published fully kinetic particle-in-cell (PIC) simulation.

The initial conditions are taken from \citet{groselj-cerri-navarro:2017} (parameters A1).
The magnetic field is ${B_x = -\delta_B \sin(2 \pi y / L)}$, ${B_y = \delta_B \sin(4 \pi x / L)}$
and ${B_z = 1 B_0}$ and the velocities are ${u_{x,s} = -\delta_u \sin(2 \pi y / L)}$,
${u_{y,s} = \delta_u \sin(2 \pi x / L)}$, ${u_{z,i} = 0}$ and
${u_{z,e} = -\frac{2 \pi}{L} \delta_B \mu_0 (2 \cos(4 \pi x / L) + \cos(2 \pi y / L))}$.
Here, the magnitude of the perturbation is given by ${\delta_u = 0.2 v_{A,0}}$ and ${\delta_B = 0.2 B_{0}}$.
In $z$-direction a current results from Faraday's law which is accounted for by the
electron velocity. Ideal MHD Ohm's law yields for the electric field
${E_x = -\delta_u B_0 \sin(2 \pi x / L)}$, ${E_y = -\delta_u B_0 \sin(2 \pi y / L)}$ and ${E_z = 0}$.
The initial density is uniform ${n_s = n_0}$ apart from a small perturbation
added to the electron density to satisfy Gauss's law. Temperatures are defined via ${T_i/T_e = 1}$
and ${\beta_i = 2 \mu_0 n_0 k_B T_i / B_0^2 = 0.1}$. Ion-electron mass ratio is set to ${m_i/m_e = 100}$
and speed of light to ${c = 18.174\,v_{A,0}}$.
The spatial resolution is $512^2$ in the Vlasov simulations (two cells per electron inertial
length) and $2048^2$ in the fluid simulation. The velocity space in the Vlasov case goes
from $-14\,v_{A,0}$ to $14\,v_{A,0}$ for the electrons and from $-1.5\,v_{A,0}$ to $1.5\,v_{A,0}$ for
the ions, each resolved by $34^3$ cells.

\begin{figure}[h!]
\includegraphics[width=\textwidth]{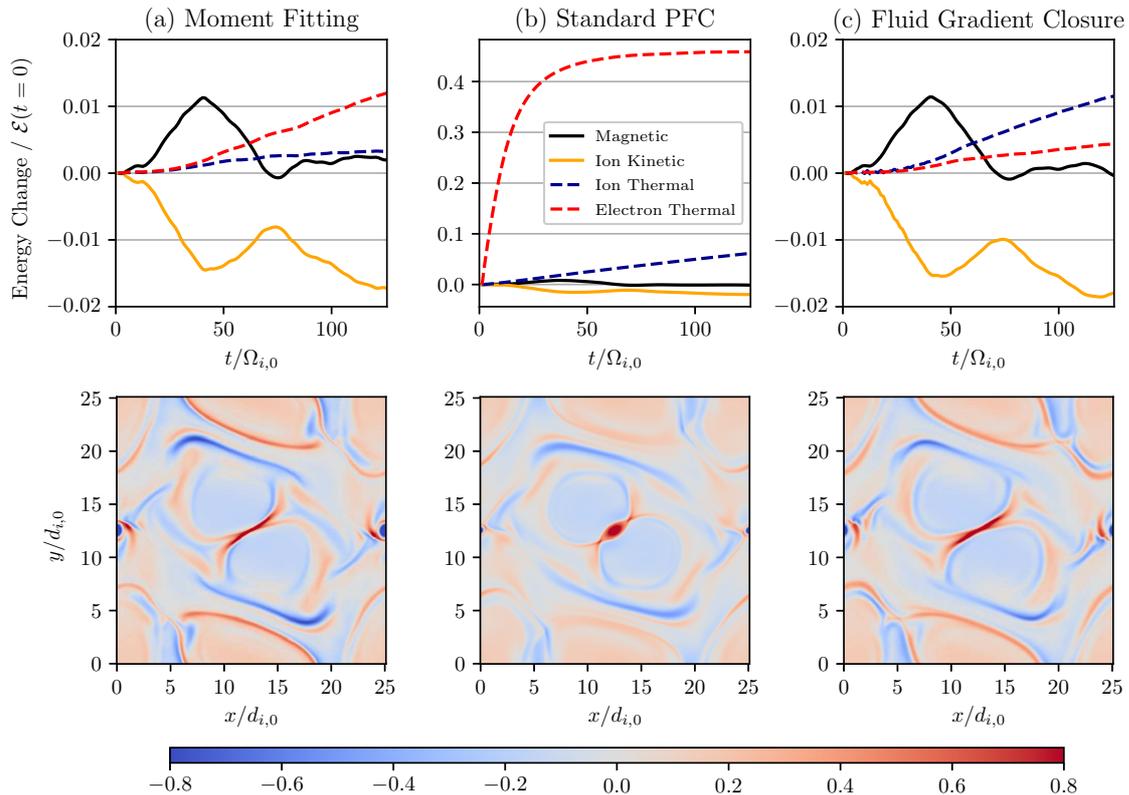}\\
\vspace{-0.5cm}
\caption{The deviation of magnetic energy, ion kinetic energy, ion thermal energy
and electron thermal energy from the initial values and
the out-of-plane current density $j_z / (q_i n_0 v_{A,0})$ at ${t=62.83\,\Omega_{i,0}}$ in Orszag-Tang turbulence.
Shown for (a) the moment fitting Vlasov solver, (b) the standard PFC Vlasov solver
and (c) the gradient closure fluid solver.
}
\label{fig:orszag_tang}\end{figure}

\begin{figure}[h!]
\includegraphics[width=\textwidth]{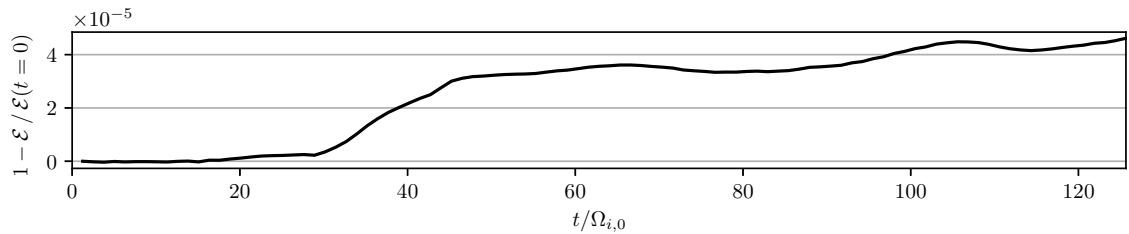}\\
\vspace{-0.5cm}
\caption{Error in total energy conservation over time in the moment
fitting Vlasov simulation of Orszag-Tang turbulence.}
\label{fig:ot_deltaf_energy}\end{figure}

Three numerical models are compared in Fig.~\ref{fig:orszag_tang}: The moment fitting Vlasov solver,
the standard PFC Vlasov solver and the fluid solver with gradient heat flux closure
(Eq.~\eqref{eq:gradient_closure}). In the upper row the distribution of energy over time is
shown whereas the lower row gives an impression of the current density's spatial
structure in out-of-plane direction. Looking at the current density, it is immediately evident
that there is good agreement between the moment fitting Vlasov model and the gradient fluid
model while the standard Vlasov solver stands out. The reason can be identified from the
energy plots which show magnetic energy ${\int \mathrm{d^3x}\ \mathbf{B}^2/(2 \mu_0)}$,
ion kinetic energy ${\int \mathrm{d^3x}\ m_i n_i \mathbf{u}_i^2/2}$ and species thermal
energies ${\int \mathrm{d^3x}\ (N/2) n_s k_B T_s}$ where $N$ is the dimensionality of velocity
space. Total energy
${\mathcal{E} = \int \mathrm{d^3x}\ \sum_s (m_s n_s \mathbf{u}_s^2/2 +  n_s k_B T_s) +
\mathbf{B}^2/(2 \mu_0) + \epsilon_0 \mathbf{E}^2/2}$ is conserved in the Vlasov-Maxwell system
with periodic boundary conditions. However, the standard Vlasov solver suffers from
substantial numerical heating so that electron
thermal energy increases to four times the initial value. In consequence, total energy reaches $1.5$ times
the initial value by the end of the simulation.
The moment fitting Vlasov solver on the other hand conserves energy well with an error smaller than
${5\cdot10^{-5}\,\mathcal{E}(t=0)}$ as shown in Fig.~\ref{fig:ot_deltaf_energy}.

In both the moment fitting Vlasov model and the fluid model ion kinetic energy is first converted
into magnetic energy. After ${t=40\,\Omega_{i,0}}$ the magnetic energy is partly converted back to
ion kinetic energy through magnetic reconnection. Over time magnetic and
kinetic energy decrease in favor of thermal energy as expected in a turbulent plasma.
In gyrokinetic turbulence \citet{kawazura-barnes-schekochihin-etal:2019} found that when the
magnetic energy is larger than the thermal energy, electrons are typically more strongly heated than ions.
This is also the case in the kinetic moment fitting simulation here, in agreement with the PIC and
gyrokinetic simulations from \citet{groselj-cerri-navarro:2017}. Generally, the evolution
of energy distribution is in excellent agreement with \citet{groselj-cerri-navarro:2017} (note
the different normalization). However, the fluid model does
not correctly predict the ratio of electron and ion thermalization. This could be improved by fine-tuning
the characteristic spatial frequencies in the gradient closure expression (Eq.~\eqref{eq:gradient_closure}),
which have direct influence on the magnitude of the heat flux and therefore on dissipation and heating.

The current density structure of both the moment fitting
kinetic model and the fluid model (Fig.~\ref{fig:orszag_tang}(d),(f)) matches that of the
PIC simulation in \citet{groselj-cerri-navarro:2017}. In the center of the domain
a current sheet has formed where magnetic field lines reconnect. The numerically heated
plasma in the standard Vlasov model features increased dissipation so that in this case
a magnetic island forms within the current sheet. The heating is less problematic for
lower mass ratios like ${m_i/m_e = 25}$ because of the smaller extent of electron velocity space and 
the resulting smaller cell sizes. In that case, however, the ion and electron scales
are not separated well leading to over- or underestimation of electron effects with
influence on the turbulence development. The geometry at the shown time has another interesting
detail which is the magnetic o-point at the left and right domain borders at
${y=L_y/2}$. The o-point is present in fully kinetic simulations and is correctly caught by the gradient fluid
model. In contrast, the -- compared to multi-fluid -- much more expensive gyrokinetic and hybrid kinetic models
do not catch this detail (see Fig.~1 of \citet{groselj-cerri-navarro:2017}).

\subsection{GEM Reconnection}\label{sec:gem}

\begin{figure}
\includegraphics[width=\textwidth]{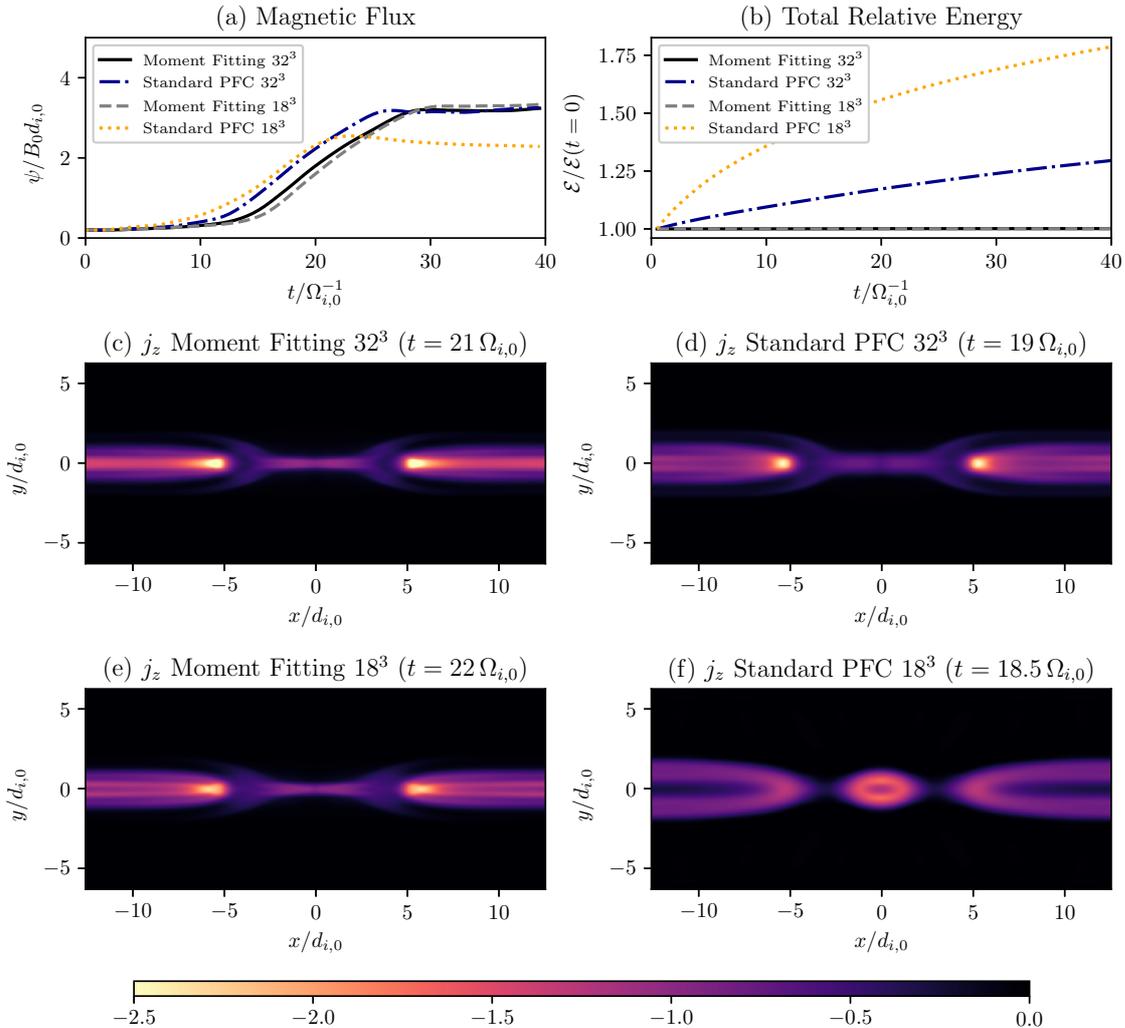}\\
\vspace{-0.5cm}
\caption{Comparison of the moment fitting and the standard PFC Vlasov solvers
at velocity space resolutions of $32^3$ and $18^3$. The out-of-plane
current density $j_z / (q_i n_0 v_{A,0})$ is shown when the magnetic flux is $\psi = 2\,B_0 d_{i,0}$.
}
\label{fig:gem}\end{figure}

One central advantage of the moment fitting Vlasov method is that
the numerical necessity for high velocity resolutions is relaxed.
To demonstrate this, we employ the GEM reconnection problem
\citep{birn-drake-shay-etal:2001} that has been studied extensively
by many authors and compare moment fitting Vlasov simulations
at low velocity space resolutions with higher-resolved moment
fitting as well as standard PFC simulations.

The initial configuration is given by a Harris equilibrium with density
${n_s = n_{0} \sech^{2}(y/\lambda) + n_{b}}$ and magnetic field
${B_{x} = \tanh(y/\lambda) B_0 + \delta B_{x}}$, ${B_{y} = \delta B_{y}}$.
The background density is ${n_{b} = 0.2\,n_0}$ and the half-width of the
current sheet is ${\lambda = 0.5\,d_{i,0}}$. The temperature is uniform and defined by
${n_{0} k_{B} (T_{e}+T_{i}) = B_{0}^{2} / (2 \mu_{0})}$, ${T_{i}/T_{e} = 5}$.
A perturbation of the magnetic field is added to initiate the reconnection process
which is given by ${\delta B_{x} = -\psi_0 \pi/L_y \cos(2 \pi x / L_x) \sin(\pi y / L_y)}$, 
${\delta B_{y} = \psi_0 2\pi/L_x \sin(2 \pi x / L_x) \cos(\pi y / L_y)}$ with ${\psi_0 = 0.1 B_0 d_{i,0}}$.
The magnetic field gradients are associated with a current density which
is distributed among electrons and ions according
to ${u_{z,i}/u_{z,e} = T_i/T_e}$. All particles contribute to the
current density without discrimination between background and sheet particles.
The reduced ion to electron mass ratio and speed of light are ${m_i / m_e = 25}$ and
${c = 20\,v_{A,0}}$, respectively.
The simulated domain is of size ${L_{x} \times L_{y} = (8\pi \times 4\pi)\,d_{i,0}}$,
here resolved by ${512 \times 256}$ cells.
It is periodic in $x$-direction, has conducting walls for fields
and reflecting walls for particles in $y$-direction and is translationally
symmetric in $z$-direction. Electron velocity space ranges from
${-12.5\,v_{A,0}}$ to ${12.5\,v_{A,0}}$ and ion velocity space from
${-5\,v_{A,0}}$ to ${5\,v_{A,0}}$.

In Fig.~\ref{fig:gem}(a) the development of the reconnected flux is compared between
the moment fitting and the standard PFC Vlasov solver at velocity space resolutions
of $32^3$ and $18^3$. Both high-resolution runs and the low-resolution moment fitting
run feature a similar slope and saturation, indicating identical reconnection physics
in the three simulations. Onset of reconnection is slightly earlier when the standard Vlasov
solver is used because of the higher electron temperature caused by numerical heating. 
The effect of large velocity space cell sizes is clearly evident in Fig.~\ref{fig:gem}(b).
While the well-resolved standard Vlasov run has an energy conservation error of 25\% at the
end of the simulation, the run with lower v-space resolution violates energy conservation
by even 75\%, accompanied by an incorrect representation of the reconnection process.
While non-positive energy conserving Vlasov schemes do not have this issue, they suffer
from numerical oscillations which at such low resolutions typically impair the reconnection
significantly and often render the simulation unstable. On the contrary, the
moment fitting Vlasov model conserves total energy for both resolutions without numerical
oscillations. Of course velocity space resolution cannot be arbitrarily low also for the
moment fitting method: The relevant physical features in the distribution function must still
be appropriately represented by the discretization and the numerical errors from the
advection scheme and the discretized integration to obtain heat flux must be sufficiently
small.

The out-of-plane current density profiles shown in Fig.~\ref{fig:gem}(c)-(f) agree as far
as the moment fitting simulations and the high-resolution standard PFC simulation are
concerned. In contrast, the low-resolution standard PFC simulation (Fig.~\ref{fig:gem}(f))
shows incorrect results. Similar to the reconnection layer in the turbulence simulation
(Fig.~\ref{fig:orszag_tang}(b)) a magnetic island forms due to the high temperature.
At the same low resolution of $18^3$, the moment fitting Vlasov method yields good
results that can even be considered more accurate than those obtained from the
standard method at $32^3$ as the current sheet is thinner and less dissipated.

\section{MMS Reconnection Events}
\subsection{Magnetotail}\label{sec:magnetotail}

\begin{figure}
\includegraphics[width=\textwidth]{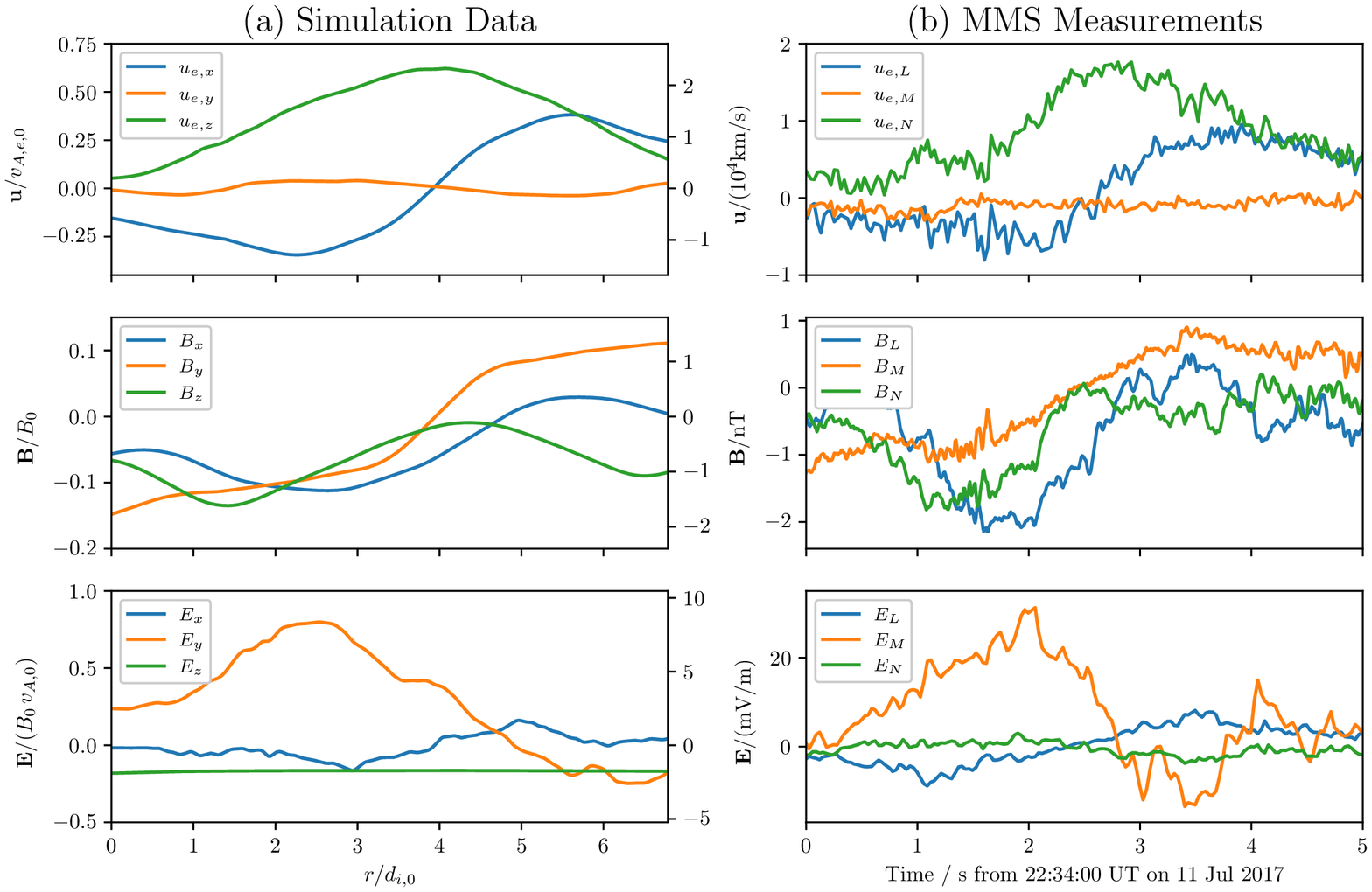}\\
\hspace{-0.9cm}\includegraphics[width=1.02\textwidth]{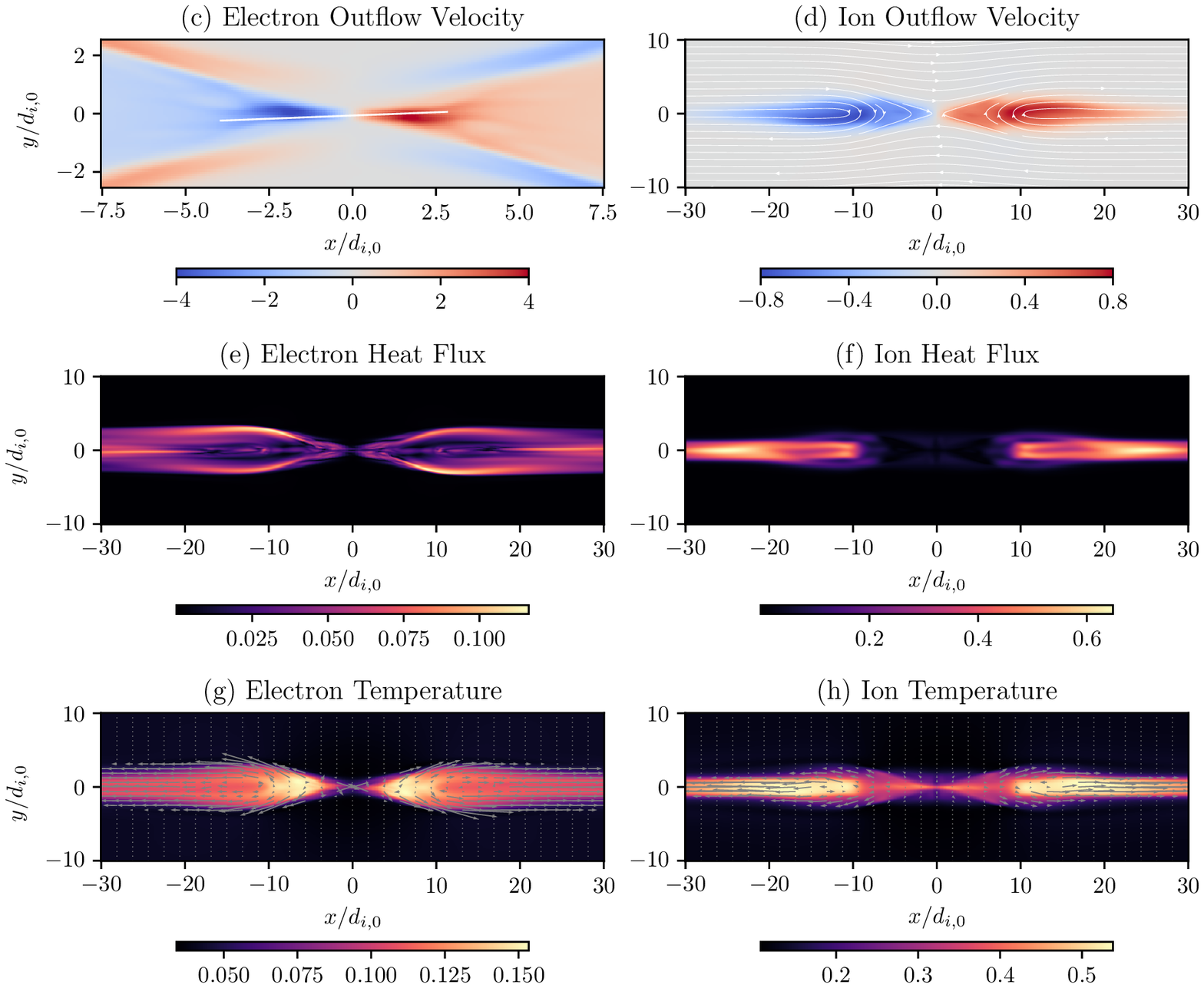}\\
\vspace{-0.25cm}
\caption{(a) Electron velocity, magnetic field and electric field along a virtual path
in the magnetotail reconnection simulation at ${t=75\,\Omega_{i}}$ and (b) as measured by MMS
and (c-h) simulation state at ${t=75\,\Omega_{i}}$.
(c) $u_{x,e} / v_{A,0}$ and the virtual path, (d) $u_{x,i} / v_{A,0}$ alongside the magnetic field lines,
(e) electron heat flux scalar $|\mathbf{q}_e| / (m_i n_0 v_{A,0}^3)$, (f) ion heat flux
scalar $|\mathbf{q}_i| / (m_i n_0 v_{A,0}^3)$, (g) $T_e/(m_i v_{A,0}^2 / k_B)$ and direction of $\mathbf{q}_e$ as arrows,
(h) $T_i/(m_i v_{A,0}^2 / k_B)$ and direction of $\mathbf{q}_i$ as arrows.
}
\label{fig:magnetotail}\end{figure}

On 11th July 2017 at 22:34 UT a reconnection event in the Earth's magnetotail
was measured by the MMS spacecraft as described in detail
by \cite{torbert-burch-phan-etal:2018}. In the two papers
by \cite{nakamura-genestreti-etal:2018} and \cite{genestreti-nakamura-nakamura-etal:2018}
physical parameters were extracted from the MMS measurements as initial
conditions for a fully kinetic PIC simulation and excellent agreement
between the simulation results and the measurement data was found as
well as accurate estimations of the reconnection rate in this magnetotail
event. Here, we perform a simulation with the same initial conditions as
in \cite{nakamura-genestreti-etal:2018} using the moment fitting Vlasov
solver to confirm their simulation results and to verify the Vlasov solver
with data from MMS measurements. The absence of noise also allows
us to analyze heat flux in our simulation and thus in the modelled
reconnection event.

The initial spatial profiles of density and magnetic field
again follow from the Harris equilibrium:
${n_s = n_{0} \sech^{2}(y/\lambda) + n_{b}}$,
${B_{x} = \tanh(y/\lambda) B_0 + \delta B_{x}}$, ${B_{y} = \delta B_{y}}$
and ${B_{z} = -B_g}$. The plasma parameters are ${n_b = n_0/3}$,
${\lambda = 0.6\,d_{i,0}}$ and guide field ${B_g = 0.03\,B_0}$. There is now
a discrimination between sheet particles (those with density ${n_{0} \sech^{2}(y/\lambda)}$)
and background particles (those with density $n_{b}$). The sheet particles
get temperatures defined by
${n_{0} k_{B} (T_{0,e}+T_{0,i}) = B_{0}^{2} / (2 \mu_{0})}$, ${T_{0,i}/T_{0,e} = 3}$ and
are responsible for the total current density. The background particles
are initially static and have temperatures ${T_{bg,s} = T_{0,s}/3}$.
Maxwellian distributions are calculated for sheet and background particles
and then added up.
Reconnection is initiated by a small Gaussian perturbation
${\delta B_{x} = -\xi \,(2 y/\lambda)\,\exp \left(-(x/(a \lambda))^2 \right) \exp\left(-(y/ \lambda)^2\right)}$
and ${\delta B_{y} = \xi \,(2 x/(a \lambda))\,\exp \left(-(x/(a \lambda))^2 \right) \exp\left(-(y/ \lambda)^2\right)}$ where ${a=L_x/L_y}$ and ${\xi = 0.01}$.
To break the symmetry we also add random noise of magnitude ${10^{-6}\,B_0}$ to $B_x$.
The domain goes from $-L_x/2$ to $L_x/2$ in $x$-direction and $-L_y/2$ to $L_y/2$
in $y$-direction and with ${L_x = 120\,d_{i,0}}$ and ${L_y = 40\,d_{i,0}}$.
Electron velocity space ranges from ${-20\,v_{A,0}}$ to ${20\,v_{A,0}}$ and ion velocity space from
${-5\,v_{A,0}}$ to ${5\,v_{A,0}}$.
We set the ion-electron mass ratio to ${m_i/m_e = 100}$ and the speed of
light to ${c = 30\,v_{A,0}}$. The resolution is ${1536 \times 512 \times 36^3}$ cells.
The simulation quantities transfer to SI units using ${n_0 = 0.09\,\mathrm{cm}^{-3}}$ and ${B_0 = 12\,\mathrm{nT}}$
\citep{nakamura-genestreti-etal:2018} so that
${d_{i,0} = 759.0\,\mathrm{km/s}}$, ${d_{e,0} = 17.71\,\mathrm{km/s}}$,
${v_{A,0} = 872.5\,\mathrm{km/s}}$, ${v_{A,e,0} = 37\,386\,\mathrm{km/s}}$
and ${E_0 = v_{A,0} B_0 = 10.47\,\mathrm{mV/m}}$.

In Fig.~\ref{fig:magnetotail} the MMS3 measurements are shown next
to the simulation data along a virtual path through the electron diffusion region,
visualized by the white line in Fig.~\ref{fig:magnetotail}c. Cell averages in
the simulation are interpolated to the path using a bivariate spline interpolator.
The simulation frame is the plasma's rest frame and the virtual path models
the movement of the plasma away from Earth through the MMS spacecraft.
The measurements are transferred from GSM coordinates to the simulation
coordinate system according to ${L = [0.9482, - 0.2551, - 0.1893]}$,
${M = [0.2651, 0.3074, 0.9139]}$, ${N = [-0.1749, -0.9168, 0.3591]}$,
which is the coordinate system obtained by \cite{genestreti-nakamura-nakamura-etal:2018}
adapted to our simulation axes. We use publicly available data from
the dual electron spectrometers \citep{pollock-moore-jacques-etal:2016}, the fluxgate magnetometer
\citep{russell-anderson-baumjohann-etal:2016} and the electric field double probe
\citep{ergun-tucker-westfall-etal:2016,lindqvist-olsson-torbert-etal:2016}.
There is excellent agreement between simulation and measurements both qualitatively
and quantitatively. Differences in magnitude, especially of the electric field,
can be attributed to the artificially reduced ion-electron mass ratio
in the simulation. The simulation also agrees very well with the much better
resolved and computationally more expensive PIC simulation in
\cite{nakamura-genestreti-etal:2018} concerning both reconnection rate
and spatial structures. Of course their simulation is
highly accurate (the resolution is better, and $m_i/m_e$ is higher) and more computational resources
have been invested than in the Vlasov simulation presented here. Nevertheless,
it should also be taken into account that we restricted our virtual path
to a straight line while they allowed fluctuations around a straight path which gives
more freedom to match the measurement data.
The Vlasov approach has advantages in the representation of the electric field which
is free of noise. This clearly shows in the good agreement between simulated
and measured electric field. The difference between the measurement of $E_N$ and
the simulation $E_z$ is due to difficulties in measuring the offset of the out-of-plane electric
field \citep{genestreti-nakamura-nakamura-etal:2018}.

Panels (c) and (d) of Fig.~\ref{fig:magnetotail} show zoomed-in views
of the electron and ion outflow velocities, respectively. Electron outflow
saturates at ${\sim 0.4\,v_{A,e,0}}$ and ion outflow has peak velocities of
${\sim 1.1\,v_{A,0}}$ at later times. The magnetic field lines shown
in panel (d) have clearly reconnected at the time. Since there is no
discrete particle noise in the Vlasov simulation, an accurate analysis of heat flux in the
simulation is possible. We plotted the magnitude of the electron heat flux vector $|\mathbf{q}_e|$
in Fig.~\ref{fig:magnetotail}(e) and its direction together with electron
temperature $T_e$ in Fig.~\ref{fig:magnetotail}(g). There are peaks in
heat flux at the separatrix borders which correspond to peaks in the electric field.
Thus, electron heat flux is dominated by energy transfer from the electric field
to the particles, one important mechanism being electron Landau damping.
Comparing the locality of heat flux with the temperature profile, it is evident that
heat flux is often located where temperature gradients are strong as for example at the separatrix
border. However, the heat flux is not necessarily along the temperature gradients
because fluctuations that are subject to Landau damping are primarily in direction of
the magnetic field which is reflected in the direction of the heat flux.
This also shows both the good potential and the deficits of the temperature gradient closure
\eqref{eq:gradient_closure} that we used for modelling Landau damping within the ten-moment multifluid
simulations of plasma turbulence in Sec.~\ref{sec:orszag_tang}. The gradient closure captures the 
location of heat flux at the temperature gradients, but the magnetic field
should be taken into account to better capture the direction of the heat flux.

In Fig.~\ref{fig:magnetotail}(f) and (h), heat flux and temperature are shown for the
ions. Ion heat flux differs significantly from electron heat flux both concerning location
and mechanism. While there is some heat flux at the separatrix boundaries, much more
is present in the outflow with a peak where the magnetic field is the strongest -- in Fig.~\ref{fig:magnetotail}
visible at ${x \approx \pm 10\,d_{i,0}}$. There, heat flux is generated through remagnetization
of the outflowing ions which start to gyrate and thus are more susceptible to wave-particle
interactions. In consequence part of their kinetic energy is converted into
thermal energy. Ion heating due to remagnetization in the ouflow
has also been measured in laboratory reconnection \citep{yamada-yoo-jara-almonte-etal:2014}.
Unexpectedly, there is a second place of
strong heat flux further downstream (starting from ${x \approx \pm 20\,d_{i,0}}$) which
is caused by fluctuations in $x$-direction, i.e.\ dominated by the $Q_{xxx}$ component of
the heat flux tensor. To weaker extent it is also present in the electrons. This second heat flux
peak is located at the head of the outflow where the outflow particles meet current sheet
and background particles that have not been accelerated in $x$-direction. One explanation
of the heat flux is a possibly increased wave activity in this region due to the outflow, accompanied
by energy transfer through wave-particle interactions. Microinstabilities related to the
different velocity distributions of the outflow particles compared to the sheet and
background particles might also cause heat flux.

\subsection{Foreshock}\label{sec:foreshock}

\begin{figure}
\centering
\includegraphics[width=\textwidth]{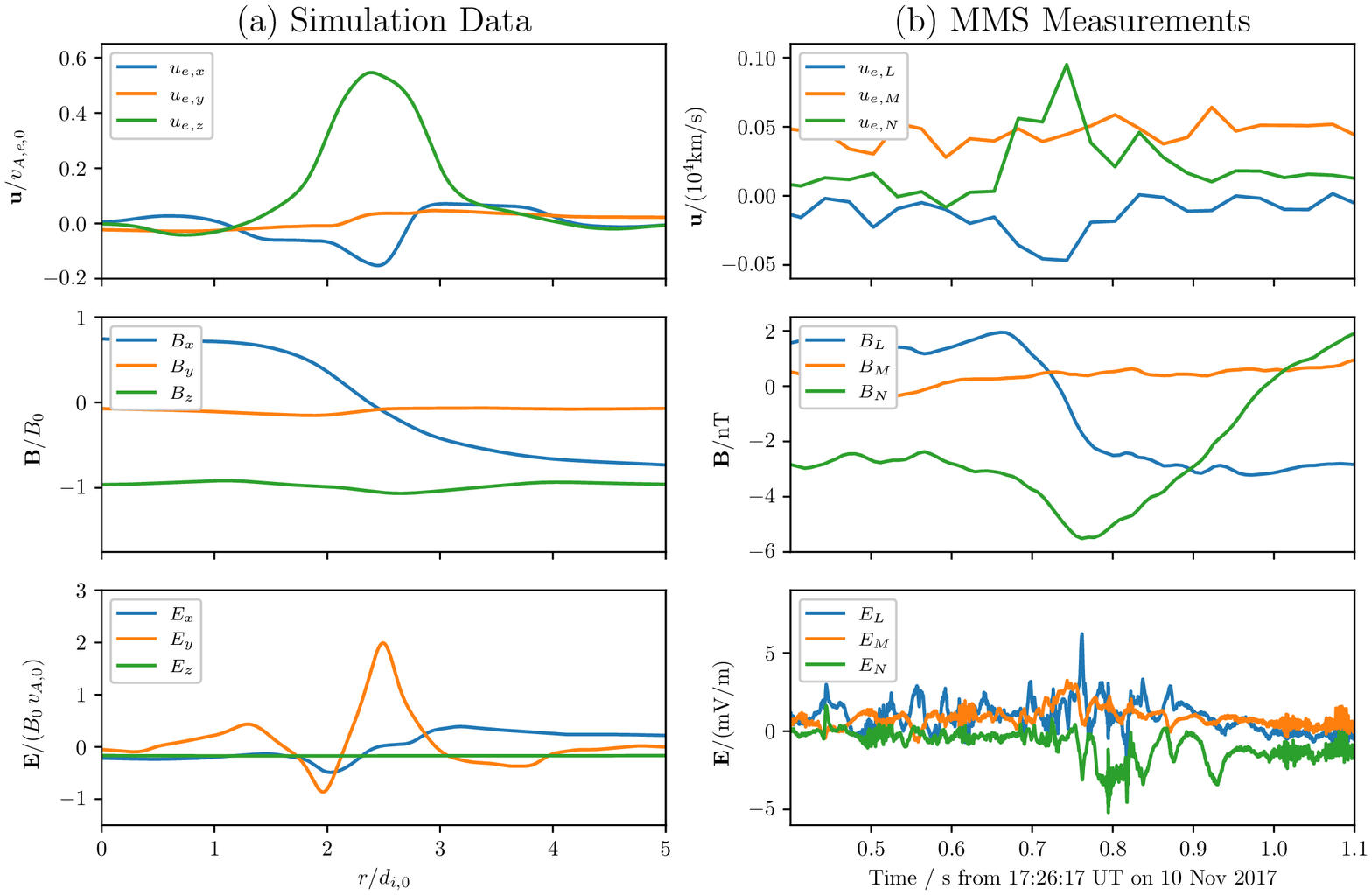}\\
\hspace{-0.33cm}\includegraphics[width=1.02\textwidth]{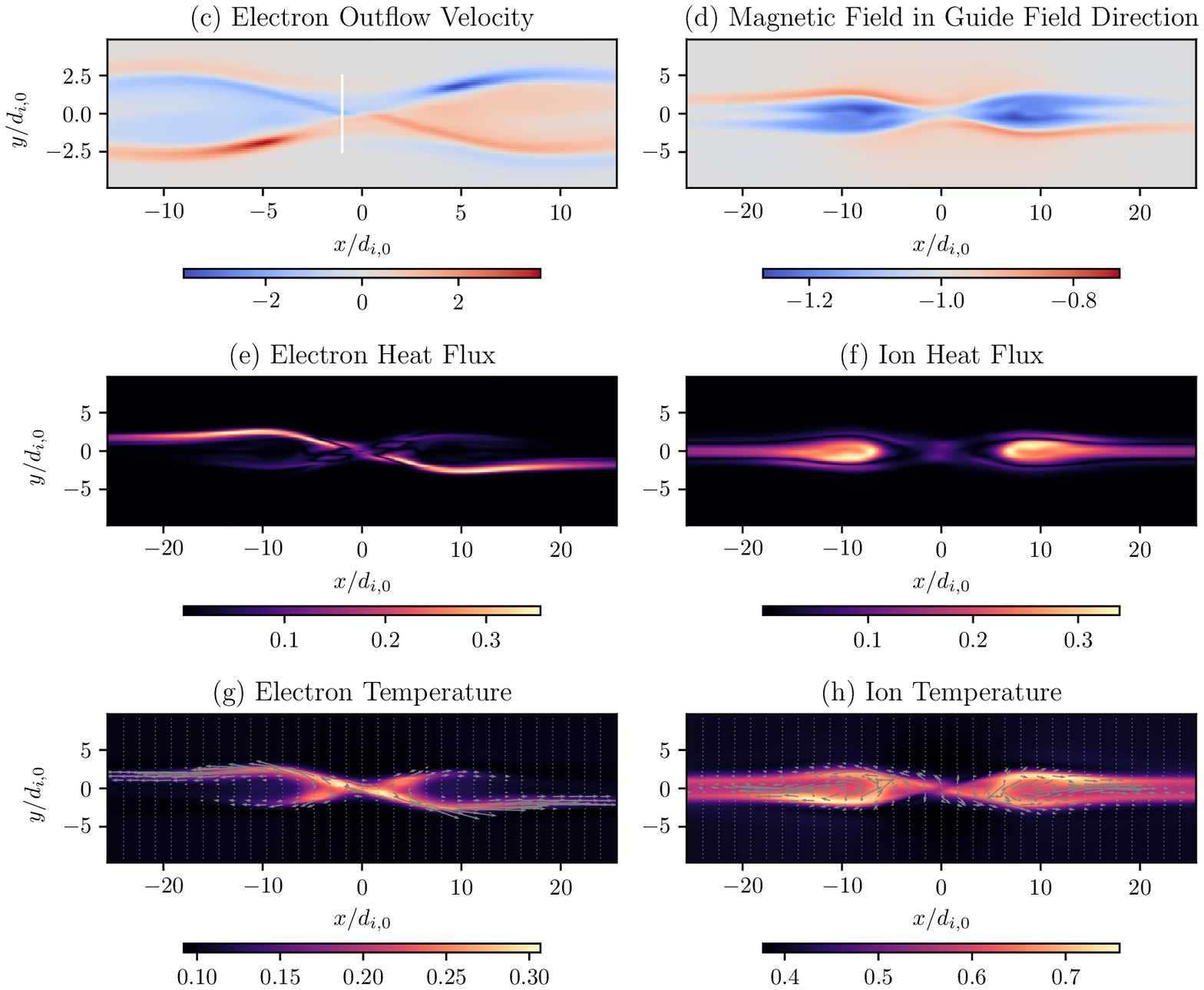}\\
\vspace{-0.25cm}
\caption{(a) Electron velocity, magnetic field and electric field along a virtual path
in the foreshock reconnection simulation at ${t=30\,\Omega_{i}}$ and (b) as measured by MMS
and (c-h) simulation state at ${t=30\,\Omega_{i}}$.
(c) $u_{x,e} / v_{A,0}$ and the virtual path, (d) $B_z / B_0$,
(e) electron heat flux scalar $|\mathbf{q}_e| / (m_i n_0 v_{A,0}^3)$, (f) ion heat flux
scalar $|\mathbf{q}_i| / (m_i n_0 v_{A,0}^3)$, (g) $T_e/(m_i v_{A,0}^2 / k_B)$ and direction of $\mathbf{q}_e$ as arrows,
(h) $T_i/(m_i v_{A,0}^2 / k_B)$ and direction of $\mathbf{q}_i$ as arrows.
}
\label{fig:foreshock}\end{figure}

In \cite{liu-lu-turner-etal:2020} two reconnection events measured by
MMS in the Earth's foreshock are reported and modelled with PIC simulations.
The event with a strong guide field that we want to discuss took place
on 10th November 2017 at 17:26:17 UT. We perform a Vlasov simulation and
compare it to measurements and PIC results.
The initial conditions are as described in Sec.~\ref{sec:magnetotail} but
with the plasma parameters that were chosen by \cite{liu-lu-turner-etal:2020}:
The guide field is now ${B_g = 1 B_0}$, the background density is ${n_{bg} = 0.2\,n_0}$,
temperature ratio is ${T_{0,i}/T_{0,i} = T_{bg,i}/T_{bg,i} = 4}$,
initial current sheet half-width is ${\lambda = 0.5\,d_{i,0}}$,
speed of light is ${c = 20\,v_{A,0}}$ and the domain is of size ${L_x = 102.4\,d_{i,0}}$,
${L_y = 25.6\,d_{i,0}}$. We use a resolution of ${1152 \times 288 \times 36^3}$
cells and ${\xi = 0.025}$ as the initial perturbation's magnitude.

A comparison between the simulation data interpolated along a
virtual path and the publicly available MMS1 data is shown in Fig.~\ref{fig:foreshock}. The measurements
are transferred from GSE coordinates to the simulation coordinate system as given by \cite{liu-lu-turner-etal:2020},
adapted to our simulation axes: ${L = [0.58, 0.24, 0.78]}$,
${M = [-0.50, 0.85, 0.11]}$, ${N = [-0.64, -0.45, 0.62)]}$. There is good qualitative
agreement between simulation and MMS data for example in the out-of-plane electron velocity $u_{z,e}$
caused by the reconnecting magnetic field and in the
electron outflow velocity $u_{x,e}$. In both simulation and measurements there is an oppositely
directed $B_x$ above and below the x-line current sheet. The guide field $B_z$ has in both cases
the same quantitative relation to the background field $B_x$. In the measured event the guide field rises strongly
after passing a local minimum which may be caused by turbulence in the foreshock reconnection
and is not seen in the two-dimensional simulation. Also, the electric fields
only fit very roughly. Quantitatively the measured electric field and the electron velocities are
rather low compared to the simulation. Quantitative agreement between model and measurements cannot
be expected -- a more precise estimate of the initial plasma parameters
and possibly a three-dimensional simulation to account for instabilities and turbulence would be
necessary. Nevertheless, the qualitative agreement suggests that the measured current sheet is indeed
due to magnetic reconnection.

The electron outflow velocity $u_{x,e}$ is shown in Fig.~\ref{fig:foreshock}(c). It differs notably
from the weak guide field case (Fig.~\ref{fig:magnetotail}(c)) and has maxima along the separatrix boundaries with low $B_z$ on
the sides where $E_x$ is in outflow direction. The outflow velocity is reduced by the guide field,
peaks at ${\sim 0.37\,v_{A,e,0}}$ and later goes back down to ${\sim 0.22\,v_{A,e,0}}$, much lower than
when the guide field is weak. On the contrary, ion outflow velocity becomes larger
(here ${\sim 1.5\,v_{A,0}}$ later in the simulation) when the guide field is strong, as has been suggested
by \cite{haggerty-shay-chasapis-etal:2018}.
For a direct comparison of the spatial structure with the PIC simulation in \cite{liu-lu-turner-etal:2020}
we have plotted the out-of-plane magnetic field in Fig.~\ref{fig:foreshock}(d). There is good agreement between
the two methods, their PIC results are overall very similar to the continuum Vlasov results here.

Electron heat flux (Fig.~\ref{fig:foreshock}(e)) in this strong guide field scenario is even more
localized at the separatrix border than in the weak guide field simulation. It is again related
to peaks of the electric field which now has a stronger preference for one side of the separatrix border
due to the stronger guide field. Electron temperature (panel (g)) has its maximum along the x-line
current sheet and is less spread out compared to the weak guide field case because heating is reduced
perpendicular to the guide field.
Ion heat flux (panel (f)) is dominated by the $Q_{zzz}$ component of the heat flux tensor. It has a peak next to
the maximum of $|B_y|$ at ${(x,y) \approx \pm (7,0)\,d_{i,0}}$ and another peak where $|B_x|$ and $|B_z|$
are both large at ${(x,y) \approx \pm (10,1)\,d_{i,0}}$. That means even when there is a strong guide field,
a rapid increase in magnetic field strength leads to a transfer of the ion kinetic energy to
thermal energy. However, it is evident in the lower magnitude of the heat flux compared to the weak guide
field case that the particles are already magnetized before entering the area where the magnetic field strength increases.
The in-plane direction of the heat flux vector shown in Fig.~\ref{fig:foreshock}(h) varies
heavily near the magnetic o-line due to the fast changing direction of the magnetic field at this place.
Ion temperature (Fig.~\ref{fig:foreshock}(h)) is the highest near the separatrix border which may be
associated with the transport of heated ions into this region along with the outflow.

\section{Conclusions}\label{sec:conclusions}

Traditional continuum Vlasov schemes have the reputation of being computationally expensive
which is due to the numerical necessity of high velocity space resolutions.
Unlike in PIC simulations, a coarse representation of velocity space may lead to
non-conservation of energy or numerical oscillations. Consequently,
continuum Vlasov simulations were primarily applied to small-scale or two-dimensional problems,
or electrons were treated non-kinetically. To address this issue, we developed
a new dual Vlasov solver which uses a standard positivity-preserving Vlasov scheme
to update the distribution function, and an energy conserving partial differential
equation solver to update velocities and temperatures. By means of moment fitting, the schemes
can work together as a positivity-preserving and energy-conserving Vlasov solver
that has good stability properties and deals well with coarse velocity space resolutions.

The new method enables us to address large-scale non-linear problems with
continuum Vlasov simulations. We performed simulations of reconnection events
measured by the MMS probe and obtained excellent agreement with measurements.
In the simulated reconnection events in the Earth's magnetosphere electron heat
flux is dominated by energy transfer from the electric field to the electrons
while ion heat flux is dominated by transfer of ion kinetic energy to thermal energy
via remagnetization. A Vlasov solver like the one
presented in this paper can compete with PIC solvers concerning computational cost.
The continuum Vlasov simulations presented in this paper agree well with published
PIC simulations, validating both methods. 
The continuum method and the PIC method have their respective strenghts and it is valuable to have different
options at hand for fully kinetic modelling of large-scale plasmas.

The next step will be spatial coupling of the Vlasov model to multi-fluid and MHD models
in order to reach global scales. The continuum Vlasov model is well-suited for smooth spatial
coupling since the noise-free distribution function is available
\citep{rieke-trost-grauer:2015,lautenbach-grauer:2018}. Using an energy conserving Vlasov solver
eases the coupling because temperature gradients
at the model interfaces due to numerical heating \citep{rieke-trost-grauer:2015}
are avoided. Ideally a plasma is represented by 
a hierarchy of models from fully-kinetic over hybrid-kinetic to multi-fluid and MHD models depending
on the plasma effects that need to be captured in the respective regions. In this hierarchy
ten-moment multifluid models can be efficiently used. In many cases they can approximate kinetic plasmas well
as we have shown in the present paper at the example of a plasma turbulence simulation. The ten-moment
multifluid model may be used either on its own or as an accurate electron model in hybrid
fluid-kinetic simulations.

There is potential to further improve the moment fitting Vlasov solver in the future. The partial
differential equation solver may be extended to not only conserve energy but also conserve momentum
(similar to \cite{amano-kirk:2013,balsara-amano-garain-etal:2016}) so that the resulting dual Vlasov solver will then
preserve positivity and conserve charge, energy and additionally momentum.
The moment fitting method is also a candidate to make low-rank Vlasov simulations \citep{kormann:2015}
conservative. Using low-rank decomposition and compression of the distribution function, much
higher velocity space resolutions (larger than $128^3$ cells) become possible.\\

\section*{Acknowledgments}

We thank Michael Abolnikov for interesting discussions and first tests of the moment fitting method.
We gratefully acknowledge the Gauss Centre for Supercomputing e.V.
(www.gauss-centre.eu) for funding this project by providing computing time
through the John von Neumann Institute for Computing (NIC) on the GCS
Supercomputer JUWELS at Jülich Supercomputing Centre (JSC).
Computations were conducted on JUWELS-booster \citep{juwels} and on the DaVinci cluster
at TP1 Plasma Research Department. F.A.~was supported by the Helmholtz Association (VH-NG-1239).
We thank the MMS team for the measurement data available at the
MMS Science Data Center (https://lasp.colorado.edu/mms/sdc/).
We used the pySPEDAS software (https://github.com/spedas/pyspedas) and the SpacePy software
(https://spacepy.github.io/) for data processing; thanks to the respective developers.\\

% \bigskip
% \noindent \textbf{Data Availability Statement}
% \input{data_statement}

\bibliography{bibliography}

\end{document}